# Sub-wavelength imaging by space dilation

Mircea Giloan and Robert Gutt

Company for Applied Informatics, Republicii nr. 107, 400489 Cluj-Napoca, Romania

**Abstract:** Two transformation-optics inspired flat lenses are used to build up an optical system capable to transpose an area surrounding the object focal point in a magnified area surrounding the image focal point. The object and image focal points of the system are placed in the free space and respectively a dilated space having a sub-unitary relative permittivity and permeability. The anisotropic and inhomogeneous media constituting the lenses enable the processing of the high spatial frequencies waves without converting them in evanescent waves. Numerical simulations show the capability of the proposed device to perform magnified discernible images of the sub-wavelength details.

Projective imaging systems based on lenses are the best option for high-speed optical microscopy. The resolution of these optical devices is limited due to the fact that sub-wavelength information from an object is carried by high spatial frequency waves which become evanescent inside of a conventional optical lens. Within the last two decades the fields of plasmonics and metamaterials provide solutions for unconventional lens designs able to process the waves of high spatial frequencies. This has provided tremendous opportunities for designing optical imaging devices with unprecedented resolution. Beginning with the seminal concept of the perfect lens [1] a number of metamaterial based lenses providing sub-wavelength resolution were studied both theoretically and experimentally [2-11].

Transformation-optics approach opened a path for designing anisotropic and inhomogeneous media, capable to manipulate not only the wave paths but also the wave vectors [12-15]. Recently, a general method was proposed for designing both converging and diverging flat lenses made of media that perform specific coordinate transformations [16]. The permittivity and permeability tensors of the transformation media constituting these unconventional lenses are retrieved from the transformation functions which alter the coordinates of the original space. A converging lens of this type provides a perfect convergence into its focal point of a light beam parallel to its optical axis. In this theoretical study, we propose an optical imaging system based on two converging transformation-optics inspired flat lenses which provides sub-wavelength resolution.

Considering the z-axis as the optical axis of the proposed system, the transformation media of the lenses used to build up the device are generated by functions which transform only the $z$ coordinate of the original space while $x$ and $y$ coordinates remain unaltered. Reference 16 describes in detail the steps leading to the retrieval of the transformation function which generates the medium of a converging lens embedded in the free space, i.e. relative permittivity and permeability are equal to unity. Following the same steps one can easily prove that for a converging lens of thickness $d$ and focal distance $\varphi$ embedded in an isotropic and homogeneous medium having the same relative permittivity and permeability, $\varepsilon=\mu=m$, the lens medium is generated by the following transformation function:

$$h(x,y) = m\left[\delta - \gamma\left(\varphi^2 + x^2 + y^2\right)^{1/2}\right] , \qquad (1)$$

where $\delta = 1 + \varphi/d$ and $\gamma = 1/d$. When $m=1$ we are in the case of a lens embedded in the free space ($\varepsilon=\mu=1$). The cases when parameter $m$ is supra-unitary ($m>1$) or sub-unitary ($m<1$) can be viewed as an additional transformation of the space like a compression or dilation, respectively. Regardless the value of parameter $m$ the optical parameters (permittivity and permeability) of the lens generated by transformation function described by eq. 1 are positive inside the area delimited by the circle given by equation:

$$x^2 + y^2 = \varphi^2 \rho(\rho+2) \ ,\qquad(2)$$

where $\rho = d/\varphi$ is the thickness to focal distance ratio of the lens.

The concept of the optical system presented in this paper is schematically represented in figure 1. The ($y$-$z$)-plane view of two identical converging flat lenses embedded in the free space (green rectangles) is depicted in figure 1(a). The lenses have a thickness to focal distance equal to four ($\rho=4$). Since the proposed device is confined to positive optical parameters of the constituent lenses the depicted area is limited on $y$-axis by the singularity point:

$$y_s = \varphi(\rho(\rho+2))^{1/2} \ ,\qquad(3)$$

derived from equation 2. The left and right lenses, depicted by red and blue rectangles, correspond to negative and positive values of $z$-coordinate and will be named the object and image lenses, respectively. The geometrical parameters, like focal distance ($\varphi$), thickness ($d$), focal point ($F$), and outside plane ($P$), are denoted for the object and image lenses with the subscripts $o$ and $i$, respectively. The curves of constant transformed $z$ coordinate are depicted by red and blue lines inside the area of the object and image lenses respectively.

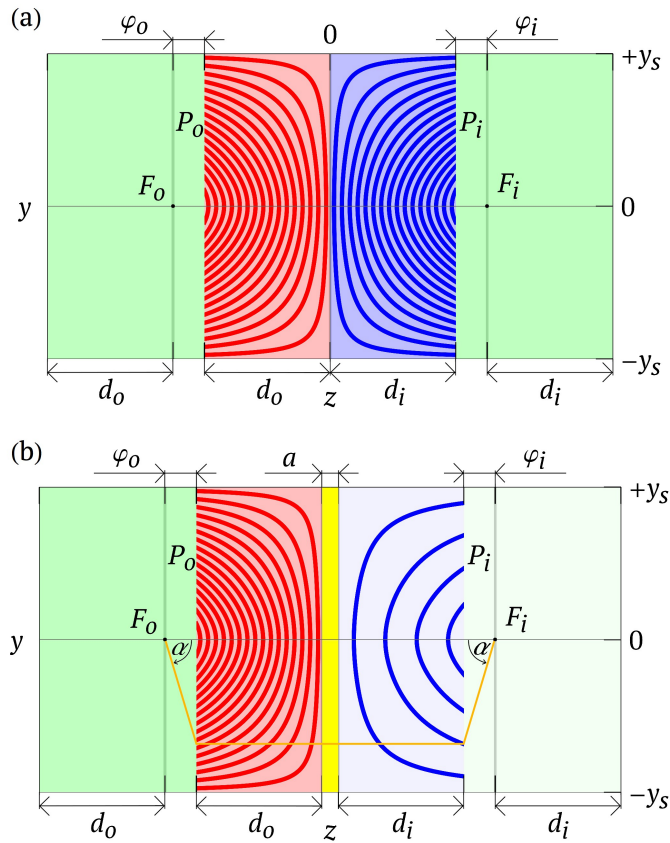

**Figure 1**. (a) ($y$-$z$)-plane view of two identical transformation-optics inspired flat lenses embedded in free space (green). The thickness, focal length and focal point of the object lens (red) are denoted b y $d_o$, $\varphi_o$ and $F_o$ respectively, while for the image (blue) lens are denoted by $d_i$, $\varphi_i$ and $F_i$ respectively. The front plane of the object lens and the back plane of the image lens are denoted by $P_o$ and $P_i$ respectively. The device span on $y$-axis is bounded by the singularity points $\pm y_s$. (b) ($y$-$z$)-

plane view of the proposed optical device. The focal points of the object (red) and image (light blue) lenses are placed in the free space (green) and a dilated space (light green), respectively. An absorption layer (yellow) of thickness *a*, is inserted between object and image lenses.

The (*y-z*)-plane view of proposed optical system is depicted in figure 1(b). The setup of the studied optical device is obtain from the setup depicted in figure 1(a) by applying two changes. The first change is that the object lens and its adjacent free space are dilated by a factor of 4. Hence, the transformation function which gives the permittivity and permeability of the object lens (light blue) is obtained from equation 1 where parameter *m* has the value 1/4 (*m*=1/4) and the space adjacent to the object lens (light green) has the permittivity and permeability equal to 1/4 ($\varepsilon=\mu=1/4$). The curves of constant transformed *z* coordinate are depicted rarefied inside the area of image lens in order to suggest the transformation by dilation of this lens area. Although, the image lens was transformed, the area of positive optical parameters (permittivity and permeability) remain the same for both lenses (see equation 2). The second change consist in an insertion of an absorbing layer (yellow) of thickness *a*, between the object and image lenses. The reason of introducing the absorption layer between lenses will be discussed later.

The key feature of the proposed optical system is that any optical ray emerging from the focal point of the object lens ($F_o$) and being incident to the outside plane of the object lens ($P_o$) will continue to propagate through the device parallel to its optical axis (*z*-axis) and when arriving at the outside plane of the image lens ($P_i$) will be deflected right through the focal point of the image lens ($F_i$). In other words, for the studied optical system the image focal point ($F_i$) is the perfect image of the object focal point ($F_o$), but in a geometrically dilated space. Due to this property the proposed optical device will transpose the object points comprised in a small vicinity around the object focal point ($F_o$) in image points comprised in a magnified vicinity around the image focal point ($F_i$).

The path of an optical ray leaving the object focal point ($F_o$) under an angle $\alpha$ with the *z*-axis and arriving into the image focal point ($F_i$) under the same angle $\alpha$ with the *z*-axis, is depicted in figure 1(b) by straight orange lines. The electromagnetic wave following this path, from $F_o$ to $F_i$, will be affected by reflection/transmission at the input ($P_o$) and output ($P_i$) interfaces of the device. In the study of this optical device, of particular interest, are the transmission coefficients at these interface planes. In reference 16 is derived the following expression of the transmission coefficient at the output interface ($P_i$) of the device as a function of $\alpha$ angle:

$$t_i = \frac{E_t}{E_i} = \frac{2}{1+\cos(\alpha)} \quad , \tag{4}$$

where $E_t$ and $E_i$ are the transmitted respective incident electric field intensity at $P_i$ interface. Following the same steps described in reference 16, a similar equation is obtained for the transmission coefficient at the input interface ($P_o$) of the device:

$$t_0 = \frac{E_t}{E_i} = \frac{2\cos(\alpha)}{1+\cos(\alpha)} \quad , \tag{5}$$

where $E_t$ and $E_i$ are the transmitted respective incident electric field intensity at $P_o$ interface. Without the absorption layer the total transmission of the device for a dipole source placed in the object focal point will be given by the following function of $\alpha$ angle:

$$t = t_0 t_i = \frac{4\cos(\alpha)}{(1+\cos(\alpha))^2} \quad , \tag{6}$$

which has a maximum value of 1 for $\alpha=0$, and it reaches the minimum value ($t_m$) for the maximum value of $\alpha$ angle ($\alpha_s$) corresponding to the singularity points ($y_s$). From equations 3 and 6 derive that the maximum value of $\alpha$ angle satisfies the following equation:

$$\cos(\alpha_s) = (\rho+1)^{-1} \tag{7}$$

and consequently the minimum value of the device total transmission is given by:

$$t_m = \frac{4(\rho+1)}{(\rho+2)^2} \tag{8}$$

where $\rho$ is the thickness to focal distance ratio of the lenses. In order to withdraw the dependence on $\alpha$ angle of the system transmission, an absorption layer was inserted between the object and image lenses. The absorption introduced by this layer should comply to the following dependence on $\alpha$ angle:

$$\tau = t_m \frac{(1+\cos(\alpha))^2}{4\cos(\alpha)} \quad, \tag{9}$$

where $t_m$ is the minimum transmission produced by the input and output interfaces of the device (see eq. 8). Taking into account the absorption effect, for a dipole source placed in the object focal point the electric field distributions at the input ($P_o$) and output ($P_i$) planes of the system are proportional with the constant factor $t_m$ which does not depends on $\alpha$ angle nor on the distance from the optical axis of the system.

The electromagnetic behavior of the proposed optical system is investigated using numerical simulations. For simplicity, the response of the designed optical device is analyzed in a two-dimensional simulation setup which reduces the computations to the field components $\{E_x, H_y, H_z\}$. As objects are used soft dipole sources generated by electric currents parallel to $x$-axis. The electric current sources are normalized in order to produce at a distance equal to the focal length an electric field intensity equal to unity. Two electric dipole sources are placed symmetrically with respect to $z$-axis and object focal point ($F_o$) at a distance equal to $3\lambda/4$, where $\lambda$ is the free space wavelength of the sources. The sources are generated coherently, i.e. in phase. In this particular two-dimensional case, the dependence on the $x$ coordinate can be eliminated from the definition of function $h$ used to retrieve the permittivity and permeability of the lenses. Hence, the transformation function is:

$$h(x,y) = m\left[\delta - \gamma(\varphi^2 + y^2)^{1/2}\right] \quad, \tag{10}$$

where $\delta = 1+\varphi/d$ and $\gamma = 1/d$ (see eq. 1), while the permittivity and permeability are given by:

$$\varepsilon = \mu = \begin{pmatrix} h(y) & 0 & 0 \\ 0 & h(y) & -zh'(y) \\ 0 & -zh'(y) & \frac{(zh'(y))^2+1}{h(y)} \end{pmatrix} \tag{11}$$

where $h'(y)$ denotes the derivative of function $h$.

The numerical simulations are performed using a two-dimensional finite-difference-time-domain (FDTD) algorithm. The values less then unity of the diagonalized tensors of permittivity and permeability are mapped with a loss-free Drude-dispersive-material model [17]. The simulation area is terminated in each direction with absorbing boundaries [18]. The simulation are performed using an orthogonal computational grid with a step resolution $\Delta=\lambda/40$ and a time step reaching the Courant limit. The component lenses of the device have $\rho=4$ and the image space is dilated by a factor $m=1/4$. The distribution of the real part of the $E_x$ field component in the simulation area of ($y$-$z$)-plane is depicted in figure 2.

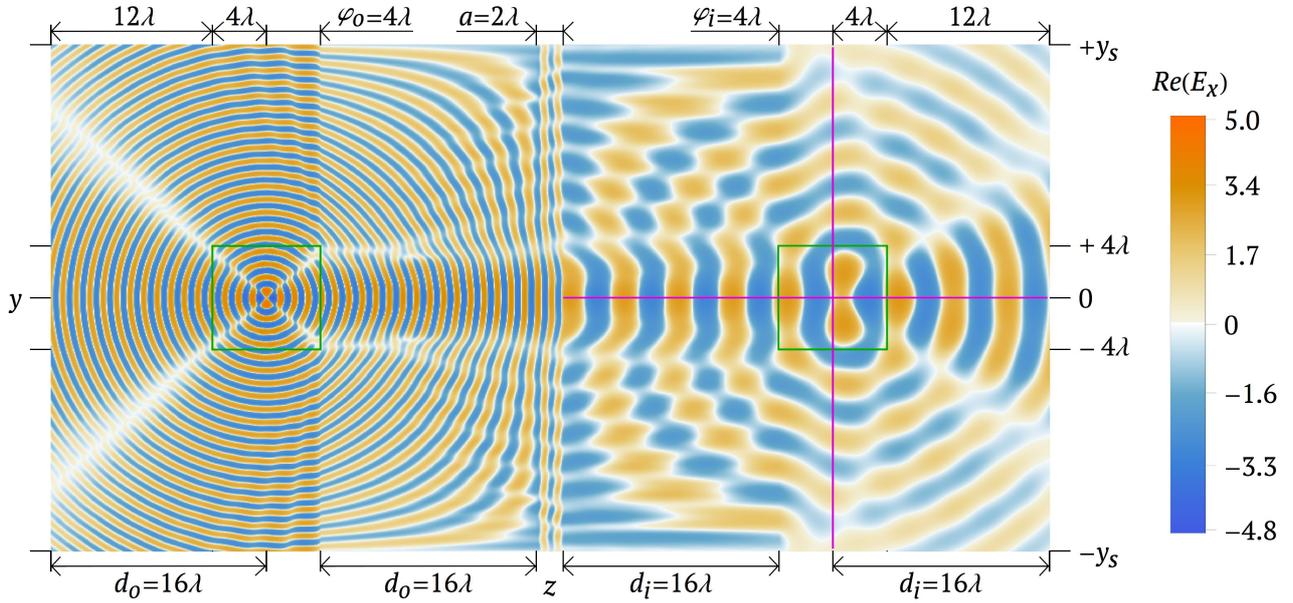

**Figure 2.** The real part of the $E_x$ component of the electric field in the simulation area of the ($y$-$z$)-plane. The thickness ($d_o$, $d_i$) and focal length ($\varphi_o$, $\varphi_i$) of the object and image lenses as well as the thickness of the absorbing layer ($a$) and other dimensions of the simulation area are expressed as multiples of free space wavelength ($\lambda$) of the dipole sources.

Two squares of side length $8\lambda$ (green) are represented in figure 2 around the object and image focal points. The electric field intensity ($E_x E_x^*$) inside these squares is depicted in figure 3. Figure 3(a) shows the intensity inside a square area surrounding the object focal point. The dipole sources are clearly observed as the points with the highest intensity. They are placed symmetrically with respect to $z$-axis and the distance between them is $3\lambda/4$.

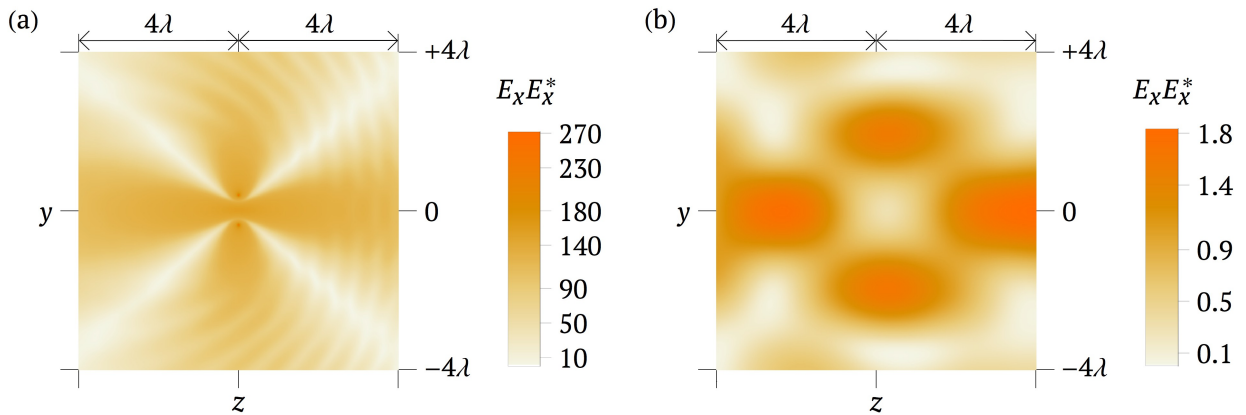

**Figure 3.** (a) Electric field intensity ($E_x E_x^*$) inside a square of side length $8\lambda$ having in its center the object focal point (see left green square in fig. 2). (b) Electric field intensity ($E_x E_x^*$) inside a square of side length $8\lambda$ having in its center the image focal point (see right green square in fig. 2).

Figure 3(b) shows the intensity inside a square area surrounding the image focal point. The image of the dipole sources are clearly observed as two area of high intensity placed symmetrically with respect to z-axis. The increased distance between the image spots compered with the distance between the dipole sources shows the magnification ability of the proposed optical device for sub-wavelength features. Two additional spots of significantly high intensity can be observed along the z-axis without disturbing the image of the dipole sources.

Figure 2 also depicts two lines (magenta) passing through the image focal point one parallel to y-axis and the other parallel to z-axis. The electric field intensity ($E_x E_x^*$) along these lines is depicted in figure 4. Figure 4(a) shows the intensity along the line parallel to y-axis passing through the image focal point. Two peaks located symmetrically with respect to the origin of y-axis represent the image of the dipole sources used in simulation. The maximum intensity of the image peaks exceeds by more then four times the maximum intensity of the side noise peaks, while the minimum intensity between image peaks almost equalize the maximum intensity of the side noise peaks. These characteristics of the electric field intensity in the image focal plane of the device clearly show the capacity of the proposed device to provide magnified discernible images of sub-wavelength features. The distance between the maximum intensity of the image peaks is about four wavelengths. Hence, the device magnification is greater then four times, since the distance between the dipole sources is set to 3λ/4.

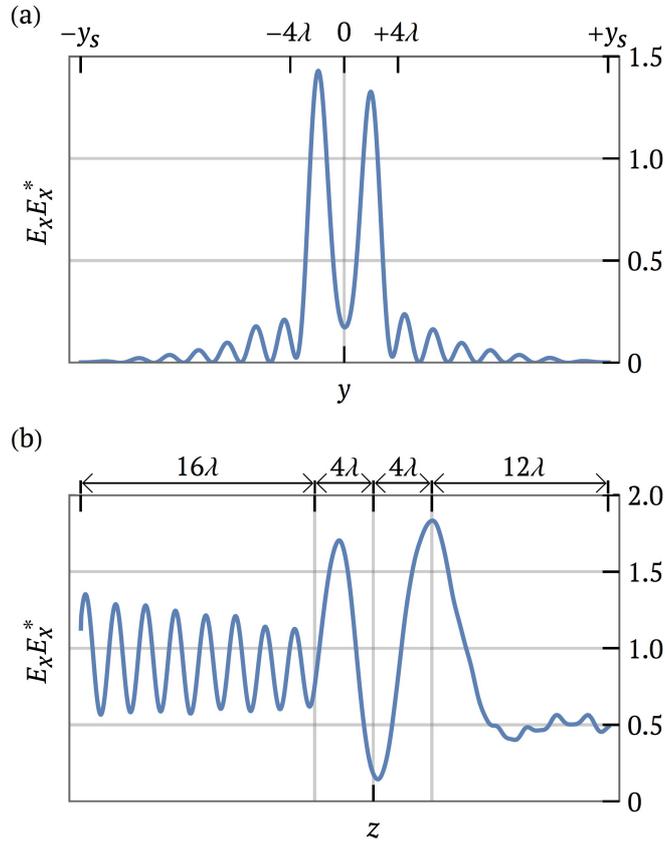

**Figure 4.** (a) Electric field intensity ($E_x E_x^*$) along the line parallel to y-axis passing through the image focal point (see vertical magenta line in fig. 2). (b) Electric field intensity ($E_x E_x^*$) along the line parallel to z-axis passing through the image focal point (see horizontal magenta line in fig. 2).

Figure 4(b) shows the intensity along the line parallel to z-axis passing through the image focal point. The intensity is plot from the input plane of the image lens to the end of simulation area. The maximum intensity of the additional peaks occurring along z-axis is higher then the maximum intensity of the image peaks. However, this graph reveals a large enough area of low electric field intensity ($E_x E_x^*$<0.5) around the position of the image focal plane which ensures no interference of these additional intensity peaks with the image peaks.

In this theoretical study, we demonstrate how two transformation optics inspired lenses can be used to build an projective optical system capable to provide magnified discernible images of objects with sub-wavelength features. The object lens has its focal point placed into the free space while the image lens has its focal point placed in a geometrically delated space having sub-unitary optical parameters. An absorption layer inserted between lenses has the aim to make the system transmission uniform. Numerical simulations show that the proposed optical device transpose sub-wavelength area around the object focal point into a magnified area around the image focal point.

**Acknowledgments:** This work was developed by the Company for Applied Informatics using infrastructure obtained through program POSCCE-A2/O2.3.2, project ID SMIS: 44023/2012, and it was partially supported by program POC-A2/221, project ID MySMIS:115641/2017.

**References:**
1. J. B. Pendry, Negative refraction makes a perfect lens. Phys. Rev. Lett. 85, 3966–3969 (2000).
2. N. Fang, H. Lee, C. Sun, and X. Zhang, Sub-diffraction-limited optical imaging with a silver superlens. Science 308, 534–537 (2005).
3. T. Taubner., D. Korobkin, Y. Urzhumov, G. Shvets, and R. Hillenbrand, Near-field microscopy through a SiC superlens. Science 313, 1595 (2006).
4. Z. Liu, H. Lee, Y. Xiong, C. Sun, and X. Zhang, Far-field optical hyperlens magnifying sub-diffraction-limited objects. Science 315, 1686–1701 (2007).
5. I. I. Smolyaninov, Y.-J. Hung, and C. C. Davis, Magnifying superlens in the visible frequency range. Science 315, 1699–1701 (2007).
6. Z. Jacob, L. V. Alekseyev, and E. Narimanov, Optical hyperlens: Far-field imaging beyond the diffraction limit. Opt. Express 14, 8247–8256 (2006).
7. Z. Liu, S. Durant, H. Lee, Y. Pikus, N. Fang, Y. Xiong, C. Sun, and X. Zhang, Far-field optical superlens. Nano Lett. 7, 403–408 (2007).
8. J. Rho, Z. Ye, Y. Xiong, X. Yin, Z. Liu, H. Choi, G. Bartal, and X. Zhang, Spherical hyperlens for two-dimensional sub-diffractional imaging at visible frequencies. Nat. Commun. 1, 143 (2010).
9. C. B. Ma and Z. W. Liu, A super resolution metalens with phase compensation mechanism. Appl. Phys. Lett. 96, 183103 (2010).
10. C. Ma and Z. Liu, Breaking the imaging symmetry in negative refraction lenses. Opt. Express 20, 2581–2586 (2012).
11. A. V. Kildishev, and E. E. Narimanov, Impedance-matched hyperlens. Opt. Lett. 32, 3432–3434 (2007).
12. M. Rahm, D. Schurig, D. A. Roberts, S. A. Cummer, D. R. Smith, and J. B. Pendry, Design of electromagnetic cloaks and concentrators using form-invariant coordinate transformations of Maxwell's equations, Photonics Nanostruct. Fundam. Appl. 6, 87 (2008).
13. W.Wang, L. Lin, J. Ma, C.Wang, J. Cui, C. Du, and X. Luo, Electromagnetic concentrators with reduced material parameters based on coordinate transformation, Opt. Express 16, 11431 (2008).
14. D. Schurig, An aberration-free lens with zero F-number, New J. Phys. 10, 115034 (2008).
15. N. Kundtz and D. R. Smith, Extreme-angle broadband metamaterial lens, Nat. Mater. 9, 129 (2010).
16. M. Giloan, Designing Devices for Wave-Vector Manipulation Using a Transformation-Optics Approach, Phys. Rev. Appl. 8, 014005 (2017).
17. N. Okada and J. B. Cole, FDTD modeling of a cloak with a nondiagonal permittivity tensor, ISRN Opt. 2012, 536209 (2012).
18. A. Oskooi and S. G. Johnson, Distinguishing correct from incorrect PML proposals and a corrected unsplit PML for anisotropic, dispersive media, J. Comput. Phys. 230, 2369 (2011).